\documentclass[10pt,aps,pra,twocolumn,showpacs,superscriptaddress]{revtex4-2}

\usepackage[colorlinks=true,linkcolor=blue,citecolor=blue,urlcolor=magenta]{hyperref}

\usepackage{amsfonts,mathrsfs,amsmath,amsthm,amssymb}
\usepackage{epsfig}
\usepackage{bm}
\usepackage{array,multirow}
\usepackage{booktabs}
\usepackage{graphicx}
\usepackage{upgreek}
\usepackage{ulem}

\begin{document}  
\title {\bf 
Statistical analysis of barren plateaus in variational quantum algorithms}

\author{Le Bin Ho} 
\thanks{Electronic address: binho@fris.tohoku.ac.jp}
\affiliation{Frontier Research Institute 
for Interdisciplinary Sciences, 
Tohoku University, Sendai 980-8578, Japan}
\affiliation{Department of Applied Physics, 
Graduate School of Engineering, 
Tohoku University, 
Sendai 980-8579, Japan}

\author{Jesus Urbaneja} 
\affiliation{Department of Mechanical and Aerospace Engineering, Tohoku University, Sendai 980-0845, Japan}

\author{Sahel Ashhab} 
\affiliation{Advanced ICT Research Institute, National Institute of Information and Communications Technology (NICT), 4-2-1, Nukui-Kitamachi, Koganei, Tokyo 184-8795, Japan}
\affiliation{Research Institute for Science and Technology, Tokyo University of Science, 1-3 Kagurazaka, Shinjuku-ku, Tokyo 162-8601, Japan}

\date{\today}

\begin{abstract}
We investigate the barren plateau (BP) phenomenon in variational quantum algorithms using a statistical approach. Using Gaussian function models, we identify three distinct types of BPs. The first type, which we called localized-dip BPs, occurs in landscapes that are mostly flat but contain a dip point where the gradient is large in a small region around the minimum. The second type, called localized-gorge BPs, which are somewhat similar to the localized-dip BPs but contain a gorge line. The third type, called
everywhere-flat BPs, appears when the entire landscape is uniformly flat with almost vanishing gradients, making optimization significantly more difficult.
After illustrating these behaviors in the Gaussian function models, we extend the analysis to the variational quantum eigensolver (VQE).  
We consider two types of ans\"atze: the hardware-efficient ansatz and the random Pauli ansatz.
For both ans\"atze, we only observe the everywhere-flat BPs.
Using our statistical approach, we searched for localized-dip and localized-gorge BPs but found no evidence of such features in the examples studied, suggesting that everywhere-flat BPs dominate in these ans\"atze. Our method effectively probes landscape features by capturing the gradient scaling across parameter space, making it a useful tool for diagnosing BPs in variational algorithms.
To mitigate BPs in the VQE, we employ a genetic algorithm (GA) to optimize the random gates generated in the ans\"atze, thereby reshaping the cost function landscape to enhance the optimization efficiency. A comparison with an unoptimized ansatz shows how the ansatz design can improve the scalability and reliability of variational quantum algorithms.
\end{abstract}
%
%
%\pacs{03.65.Ta, 03.65.Aa, 02.50.-r, 03.67.Ac }
\maketitle

\section{Introduction}
Variational quantum algorithms (VQAs) are central to the development of quantum computing on near-term hardware. By combining parameterized quantum circuits with classical optimization, VQAs offer a flexible framework for tackling a wide range of problems in quantum chemistry, machine learning, and many-body physics \cite{Cerezo2021}. Their compatibility with noisy intermediate-scale quantum (NISQ) devices makes them particularly attractive for practical implementation in the near future.

However, a major limitation in scaling VQAs is the barren plateau (BP) phenomenon \cite{McClean2018}, which refers to the exponential vanishing of cost function gradients with increasing system complexity. This phenomenon occurs when the cost function landscape becomes extremely flat, hindering numerical optimizers from finding the optimal direction for improvement with gradient-based methods. 

BPs are influenced by the structure of the parameterized quantum circuits (ans\"atze) and the choice of cost functions, which include random $t$-design ansatz \cite{McClean2018, Cerezo2021_1}, ansatz expressibility \cite{PRXQuantum.3.010313, Larocca2022diagnosingbarren}, global cost functions and cost function concentration \cite{Uvarov_2021, Cerezo2021_1, Arrasmith_2022}, noise \cite{Wang2021} and entangled ansatz \cite{PRXQuantum.2.040316}. Recently, these sources of BPs have been unified under a common framework based on Lie algebra theory \cite{Ragone2024, Fontana2024}, which provides a deeper understanding of their origin.

\begin{figure*}[t]
    \centering
    \includegraphics[width=\textwidth]{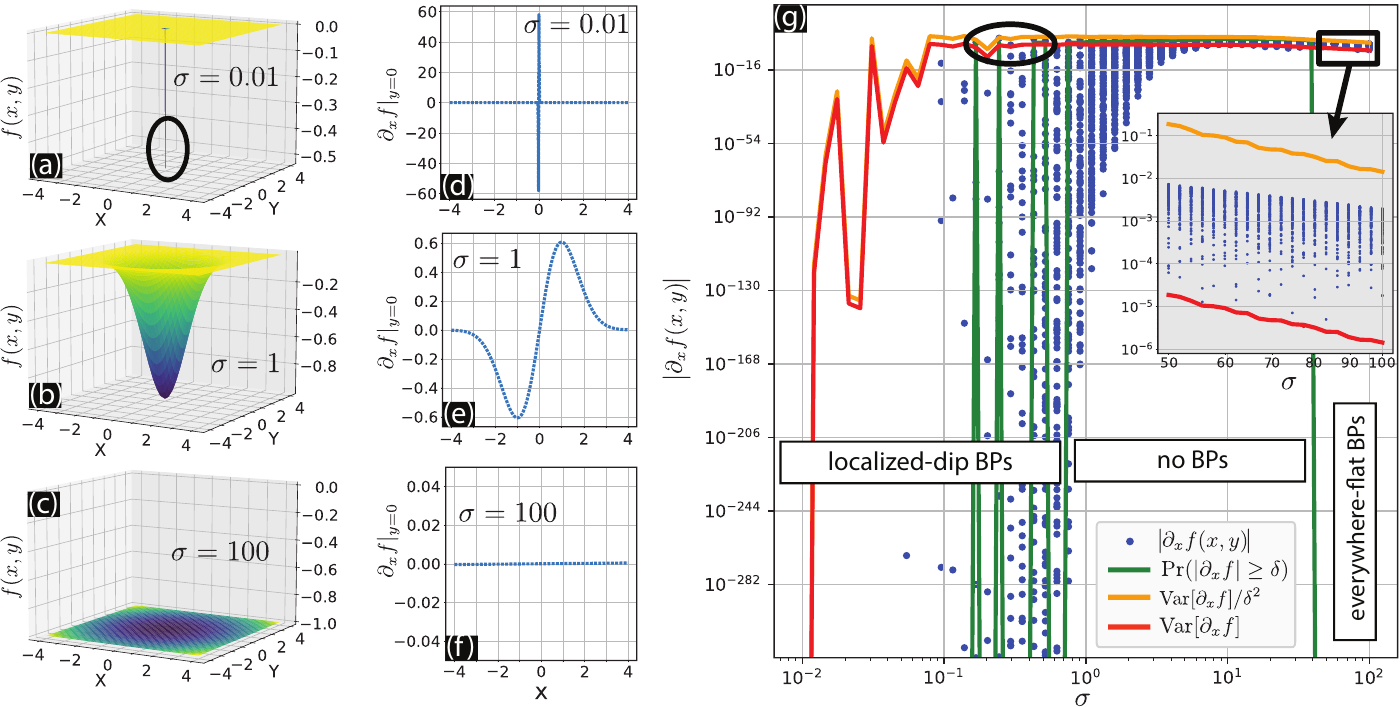}
    \caption{(a–c) 3D plots of the isotropic Gaussian function $f(x,y)$ in Eq.~\eqref{eq:fxy_G} for different values of \(\sigma\). The landscapes in (a) and (c) are mostly flat, except for a deep peak in (a), highlighted by the black oval.  
    (d–f) Derivatives of the Gaussian function $\partial_x f|_{y=0}$ plotted against \(x\) with \(y = 0\).  
    (g) Plot of the statistical distribution of the derivatives $|\partial_xf|$, the threshold probability \({\rm Pr}(|\partial_x f| \geq \delta)\), the Chebyshev bound \({\rm Var}[\partial_x f]/\delta^2\) for \(\delta = 0.01\), and corresponding variance ${\rm Var}[\partial_x f]$. The probability \({\rm Pr}(|\partial_x f| \geq \delta)\) divides the derivatives into three regions, with BPs appearing in the first and third regions. In the localized-dip plateau region, certain parameter sets \((x, y)\) avoid BPs by falling within the dip point, marked by the black oval in PBs.
    Inset:  A zoom-in of the everywhere-flat region with $\sigma \in [50, 100]$, illustrating the exponential decay of the derivatives, while showing that the distribution's width on a logarithmic scale depends only weakly on $\sigma$.
    }
    \label{fig:1}
\end{figure*}

To mitigate BPs, several strategies have been proposed, including engineered parameter initialization \cite{Liu_2023,Grant2019initialization, Volkoff_2021}, low-entanglement circuit initializations \cite{PhysRevResearch.3.033090}, shallow circuit designs \cite{Cerezo2021_1}, 
classical shadows \cite{PRXQuantum.3.020365}, local cost functions \cite{Cerezo2021_1}, transferability of smooth solutions \cite{PhysRevA.106.L060401}, pre-training \cite{PhysRevA.106.042433,Dborin_2022,LIU2022128169,kieferova2021quantumgenerativetrainingusing}, and engineered dissipation \cite{Sannia2024}. These approaches aim to preserve gradients and enhance trainability.

Previous studies have primarily examined BPs in specific circuit architectures or initialization schemes, using analytical tools and numerical simulations \cite{McClean2018, Cerezo2021_1,PRXQuantum.3.010313, Larocca2022diagnosingbarren,Uvarov_2021, Cerezo2021_1, Arrasmith_2022,Wang2021,PRXQuantum.2.040316,Ragone2024, Fontana2024}. 
While these works identify when and where BPs occur, they do not address the possible differences between different types of BPs. 
Understanding the statistical properties of BPs can provide insight into different possible scenarios and can help guide the development of strategies to effectively mitigate the deleterious effects of BPs.

In this work, we introduce a statistical framework to explain the origins of BPs by analyzing the cost function landscape and its distribution in parameter space. We identify three distinct types: one where the landscape is flat almost everywhere but contains a sharp dip with a steep slope, 
one that contains a sharp gorge, and one where the landscape is uniformly smooth and flat. 

We analyze gradient behavior using example models of Gaussian functions, which exhibits these types of BPs. We then extend the analysis to VQE models with the hardware-efficient ansatz (HEA) and random Pauli ansatz (RPA). We find that these ans\"atze are dominated by everywhere-flat plateaus. 

We then propose a mitigation strategy using a genetic algorithm \cite{Hai_2024} to eliminate BPs by optimizing the random gates generated in the ansatz, thereby reshaping the cost function landscape and enhancing its gradients for more effective optimization. This study presents a statistical perspective on BPs and provides a practical solution to improve the scalability and reliability of VQAs.

The rest of the paper is organized as follows. In Section~\ref{sec2}, we present the statistical analysis of BPs in the Gaussian function models, and in Section~\ref{sec3}, we focus on VQE. Section~\ref{sec4} details the proposed mitigation strategy and its application to quantum circuits. We conclude in Section~\ref{sec5} with a discussion of implications and future directions.

\begin{figure*}[t]
    \centering
    \includegraphics[width=0.7\textwidth]{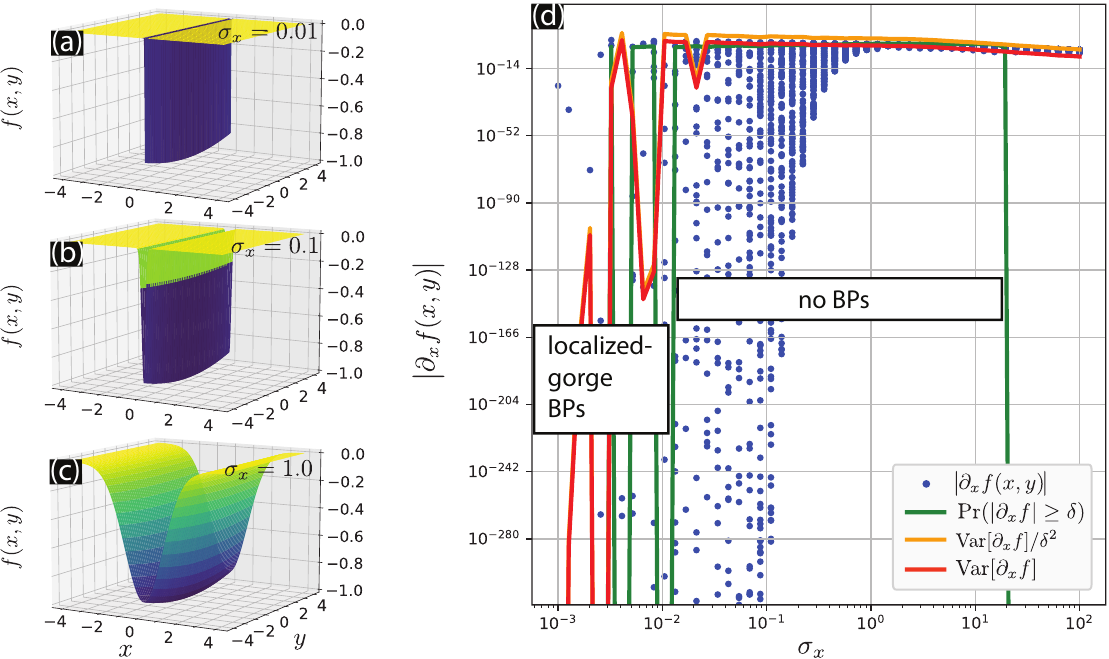}
    \caption{
(a–c) 3D plots of the anisotropic Gaussian function $f(x,y)$ in Eq.~\eqref{eq:fxy_g} for various values of $\sigma_x$, with $\sigma_y = 10$. In (a), the landscape is mostly flat except for a deep gorge along $x = 0$, in contrast to the localized dip observed in Fig.~\ref{fig:1}(a).
(d) Statistical analysis of the derivatives $|\partial_x f|$, including the distribution, the threshold probability ${\rm Pr}(|\partial_x f| \geq \delta)$, the Chebyshev bound ${\rm Var}[\partial_x f]/\delta^2$ with $\delta = 0.01$, and the variance ${\rm Var}[\partial_x f]$, following the same procedure as in Fig.~\ref{fig:1}(g). All other parameters are the same as those in Fig.~\ref{fig:1}.
}
    \label{fig:2}
\end{figure*}

\section{BPs in model functions}\label{sec2}

In this section, we investigate the statistical behavior under different geometrical plateau profiles, specifically narrow dip/gorge and flat structures, using simple 2D Gaussian functions as examples. %This analysis introduces some useful tools for evaluating the practical applicability and limitations of statistical methods in VQAs. 
We use two Gaussian functions: an isotropic 2D Gaussian with equal standard deviations in both directions, and an anisotropic 2D Gaussian with different standard deviations along $x$ and $y$.

\subsection{Isotropic 2D Gaussian}
We begin by examining an isotropic 2D Gaussian function as a cost function and its derivative with respect to \(x\)
\begin{align}\label{eq:fxy_G}
    f(x, y) = - \exp\left(-\frac{x^2 + y^2}{2\sigma^2}\right),
\end{align}
where \(\sigma\) is the standard deviation.
The derivative with respect to $x$ gives
\begin{align}
\frac{\partial f (x,y)}{\partial x} = \frac{x}{\sigma^2} \exp\left(-\frac{x^2+y^2}{2\sigma^2}\right).
\end{align}

Figure~\ref{fig:1}(a–c) shows \(f(x, y)\) for three values of \(\sigma\). For very small \(\sigma\), e.g., \(\sigma = 0.01\), it is flat except for a sharp dip centered at  \(x = y = 0\), marked by the black oval in Fig.~\ref{fig:1}(a). The corresponding derivative \(\partial f/\partial x\big|_{y=0}\), shown in Fig.~\ref{fig:1}(d), is almost zero everywhere except for a high-peak-deep-valley structure at the dip point. This behavior indicates a BP, even though this behavior does not occur at the dip, as will be discussed later. As \(\sigma\) increases to moderate values (figure b), \(f(x, y)\) forms a more defined shape, and \(\partial f/\partial x\big|_{y=0}\) in Fig.~\ref{fig:1}(e) remains non-zero across a wider range of \(x\), indicating the absence of BPs. When \(\sigma\) becomes large, \(f(x, y)\) flattens further as shown in Fig.~\ref{fig:1}(c), and \(\partial f/\partial x\big|_{y=0}\) vanishes everywhere as shown in Fig.~\ref{fig:1}(f), which is a characteristic feature of BPs.

The first and third cases both exhibit BPs, but the BPs in the two cases arise from different causes. For very small $\sigma$, the plateau is correlated with the development of a sharp dip at the origin, referred to as a localized-dip BP. In contrast, for large $\sigma$, the landscape is uniformly flat with no guiding features, leading to an everywhere-flat BP.

To further explore the different types of BPs, we generate random \(x, y\) values uniformly distributed in the range \([-20, 20]\) and examine the distribution of \(|\partial f/\partial x|\), as shown in Fig.~\ref{fig:1}(g). The results show that for both very small and very large \(\sigma\), most derivatives are close to zero. This confirms that the gradients vanish across these domains, indicating BPs. 
For localized-dip BPs, the derivatives become extremely small for most of the parameter space, while rare large derivatives continue to exist. In contrast, for everywhere-flat BPs, the derivatives do not decrease as dramatically, because they remain on the order of the ratio between range of cost function values and the range of variable parameter values.

To quantify this behavior, we apply Chebyshev’s inequality, a statistical tool that bounds the probability of a variable deviating from its mean
\begin{align}\label{eq:Che_in}
    {\rm Pr}\Big(\Big| \partial_x f - \langle\partial_x f\rangle \Big| \geq \delta \Big) \leq \frac{{\rm Var}[\partial_x f]}{\delta^2},
\end{align}
where \(\langle \partial_x f \rangle\) is the mean, \({\rm Var}[\partial_x f]\) is the variance, and \(\delta > 0\) is a chosen threshold. In many cases, the average derivative \(\langle \partial_x f \rangle\) is zero due to the symmetry of the landscape.

This inequality shows that when the variance is small, the probability of observing large gradients becomes negligible. Thus, Chebyshev’s inequality provides a rigorous way to detect BPs: if the variance of the gradient is low, significant deviations from the mean are unlikely, indicating an almost flat landscape.

When BPs occur, the probability \({\rm Pr}(|\partial_x f| \geq \delta)\) becomes negligible, as described by Chebyshev’s inequality. This is the statistical explanation for the occurrence of flat gradients in the optimization landscape.
Figure~\ref{fig:1}(g) plots \({\rm Pr}(|\partial_x f| \geq \delta)\) (green) and its Chebyshev bound \({\rm Var}[\partial_x f]/\delta^2\) (orange) for \(\delta = 10^{-2}\). 
In the central region, the inequality is violated, that is, the observed probability exceeds the bound, indicating no BP. In contrast, on the left and right sides, the probability drops to zero and satisfies the inequality, signaling a BP.
However, the inequality may become loose when \({\rm Var}[\partial_x f]/\delta^2 > 1\), since \({\rm Pr}(|\partial_x f| \geq \delta) \leq 1\) trivially holds. Therefore, while Chebyshev’s bound helps identify BPs, its interpretability depends on how tight the bound is in each region.

Our statistical analysis suggests that in the presence of local-dip BPs, there is still a chance to reach the global minimum. For example, in the left region of Fig.~\ref{fig:1}(g), certain values of $x$ near or at the dip point (marked by the black oval) yield non-vanishing gradients. This indicates that local dips may provide an opportunity for optimization. This enables statistical analysis to identify such regions prior to training, rather than using random initialization. In contrast, the right region exhibits an everywhere-flat plateau, where $|\partial f/\partial x|$ decreases exponentially across the entire domain, making optimization more challenging. 
The inset focuses on the everywhere-flat region for $\sigma \in [50, 100]$, demonstrating the exponential decay of the derivatives and revealing that the distribution's width on a logarithmic scale depends weakly on $\sigma$.

\subsection{Anisotropic 2D Gaussian}

Next, we consider an anisotropic 2D Gaussian model, where $\sigma$ differs for different parameters, i.e.,
\begin{align}\label{eq:fxy_g}
    f(x, y) = - \exp\left(-\frac{x^2}{2\sigma_x^2} - \frac{y^2}{2\sigma_y^2}\right).
\end{align}

We examine the case where $\sigma_y$ is fixed (e.g., $\sigma_y = 10$, as shown in Fig.~\ref{fig:2}). The overall landscape resembles that of the isotropic case. However, as $\sigma_x$ decreases, the central well progressively contracts along the $x$-axis, eventually forming a deep gorge, which is clearly visible in Fig.~\ref{fig:2}(a) for $\sigma_x = 0.01$. We refer to this configuration as the localized-gorge BPs.

To distinguish between the localized-dip and localized-gorge BPs, we analyze the statistical behavior in Fig.~\ref{fig:2}(d), following the same approach as in Fig.~\ref{fig:1}(g). Overall, the statistical profiles in both cases appear qualitatively similar.
However, the no-plateau region (middle region) is broader than in the previous case and extends further toward smaller $\sigma_x$. Furthermore, the localized-gorge BPs emerge only when $\sigma_x$ becomes significantly smaller than in the localized-dip scenario, marking a key distinction between the two cases.

This observation underscores an important point: despite the similarity in statistics along the $x$-direction, the overall landscape is less confined owing to the lack of contraction along the $y$-axis. Consequently, the localized-gorge BPs arise in an extreme parameter regime and exert a weaker influence on the training process, as they occupy a narrower region of the parameter space and have less impact on gradient suppression.

\begin{figure}[t]
\centering
\includegraphics[width=\columnwidth]{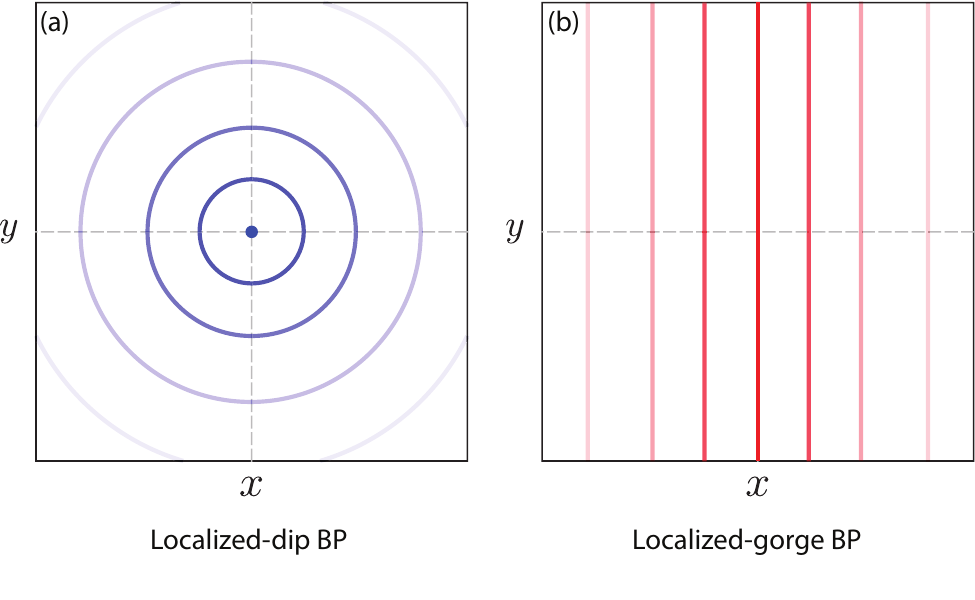}
\caption{
Illustration of the localized-dip BP (a) and localized-gorge BP (b).
(a) The central well contracts uniformly in all directions, forming a sharp dip at $x = y = 0$.
(b) The central well contracts along the $x$-axis, resulting in a sharp gorge at $x=0$.
} 
\label{fig:3}
\end{figure}

For further comparison, Fig.~\ref{fig:3} highlights these two types of BPs. In the localized-dip case, the central well contracts uniformly in all directions, forming a sharp dip at $x = y = 0$. In contrast, the localized-gorge case exhibits contraction along the $x$-axis, resulting in a narrow, elongated gorge at $x = 0$.

\section{BPs in VQAs}\label{sec3}
VQAs are hybrid algorithms that combine quantum circuits with classical optimization techniques to solve complex problems. % in many-body physics, quantum chemistry, combinatorial optimization, and more. 
Since VQAs are based on a heuristic approach, they can be applied to any computational problem that can be posed as an optimization problem. The performance of any specific VQA then depends on the quality of the quantum circuit structure and parameter rules used by the algorithm.
In these algorithms, a system evolves under a parameterized circuit \( U(\boldsymbol{\theta}) \), where \(\boldsymbol{\theta}\) are trainable parameters. A classical optimization algorithm then iteratively updates these parameters to minimize a cost function \(C(\boldsymbol{\theta})\), driving the system toward an optimal solution.

Let \( \partial_{\theta_i} C \equiv \partial C / \partial \theta_i \) denote the derivative of the cost function \( C(\boldsymbol{\theta}) \) with respect to a parameter \( \theta_i \). 
In a problem that is characterized by the presence of BPs, the variance of this gradient scales as \( \text{Var}(\partial_{\theta_i} C) \propto e^{-N} \), where \( N \) is the system size (e.g., the number of qubits) \cite{McClean2018}. This exponential decay limits the effectiveness of gradient-based optimization.

Hereafter, we analyze BPs in VQAs using the statistical approach introduced in the previous section.

\subsection{Ans\"atze}
\begin{figure}[t]
\centering
\includegraphics[width=\columnwidth]{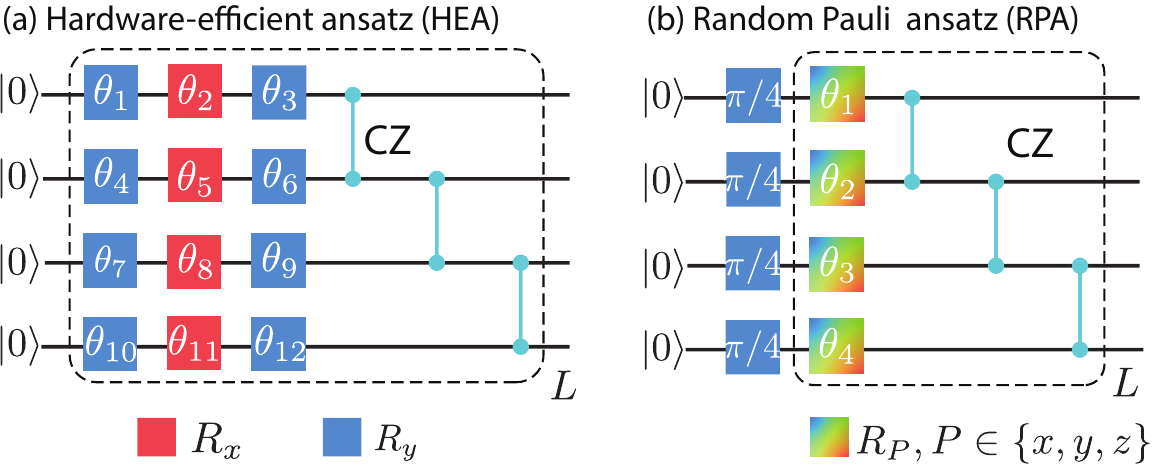}
\caption{Illustration of a parameterized quantum circuit with \( N = 4 \) qubits.  
(a) Hardware-efficient ansatz (HEA): The circuit starts from the initial state \( |\bm{0}\rangle \), followed by a sequence of \( R_y \)-\( R_x \)-\( R_y \) rotations on each qubit. Controlled-Z (CZ) gates are then applied between adjacent qubits in a linear topology. This structure is repeated over \( L \) layers.  
(b) Random Pauli ansatz (RPA): The circuit begins with a fixed layer of \( R_Y(\pi/4) \) gates applied to all qubits, followed by \( L \) layers of randomly selected single-qubit Pauli rotations and CZ gates.}
\label{fig:4}
\end{figure}

We examine a benchmark VQA \cite{Peruzzo2014} using two ans\"atze: the hardware-efficient ansatz (HEA) \cite{Kandala2017} and the random Pauli ansatz (RPA) \cite{McClean2018}, as illustrated in Fig.~\ref{fig:4}.
The HEA is composed of a sequence of parameterized quantum gates, where the unitary operator \( U(\boldsymbol{\theta}) \) is typically structured as
\begin{align}
U(\boldsymbol{\theta}) = \prod_{\ell=1}^{L} \left( U_{\text{ent}}^{(\ell)} \cdot U_{\text{rot}}^{(\ell)}(\boldsymbol{\theta}^{(\ell)}) \right),
\end{align}
where \( L \) is the number of layers, \( U_{\text{rot}}^{(\ell)} \) includes parameterized single-qubit rotations, 
i.e., $U_{\rm rot}(\bm \theta^{(\ell)}) = \bigotimes_{j=1}^N R_y(\theta^{(\ell)}_{j,3}) R_x(\theta^{(\ell)}_{j,2})
R_y(\theta^{(l)}_{j,1})$, and \( U_{\text{ent}}^{(\ell)} \) represents entangling gates such as a fixed combination of CZ gates applied to the qubits. Note that the different CZ gates commute and can be implemented simultaneously.
For  simplicity, we rewrite $\theta^{(\ell = 1)}_{1,1}$ as $\theta_1$; $\theta^{(\ell = 1)}_{1,2}$ as $\theta_2$, and so on.
The total number of parameters is $d = 3NL$.

\begin{figure*}[t]
\centering
\includegraphics[width=0.7\linewidth]{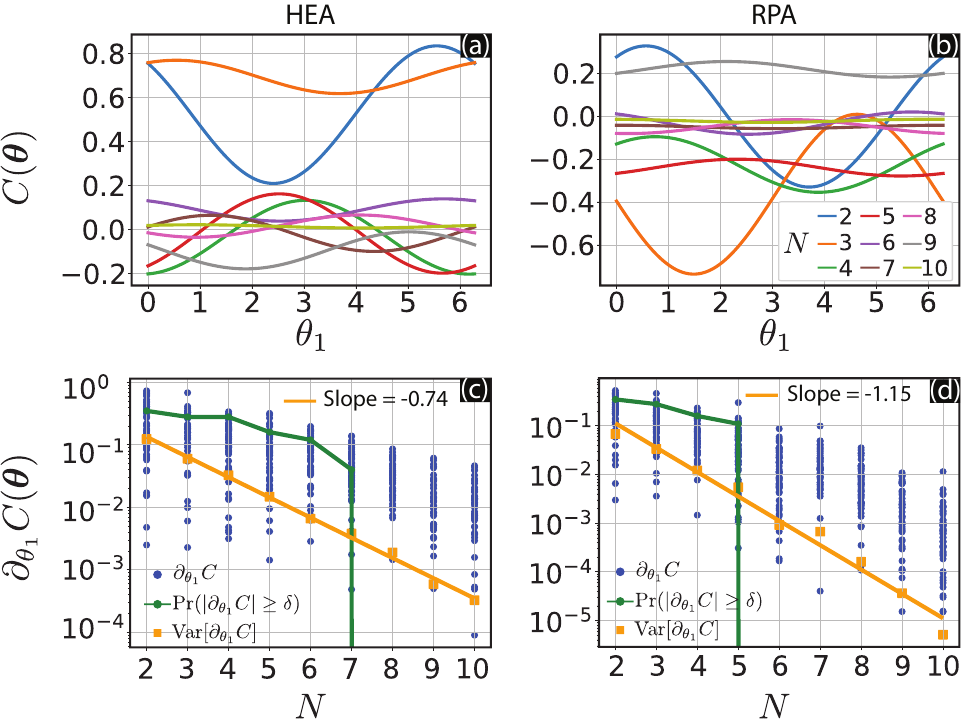}
\caption{Numerical analysis of BPs for HEA and RPA with \(L = 20\).  
(a, b) Cost function \(C(\boldsymbol{\theta})\) as a function of \(\theta_1\) for various system sizes \(N\), with all other parameters randomly fixed, shown for HEA (a) and RPA (b).  
(c, d) Derivative distributions \(\partial_{\theta_1} C\) under random sampling of \(\bm\theta \in [0, 2\pi]\), including the probability \(\mathrm{Pr}(|\partial_{\theta_1} C| \geq \delta)\) at \(\delta = 10^{-1}\) and the variance \({\rm Var}[\partial_{\theta_1} C]\), corresponding to the settings in (a) and (b), respectively and their fitting lines.}
\label{fig:5}
\end{figure*}

The RPA begins with a fixed layer of \( R_Y(\pi/4) \) gates applied to all qubits, followed by a layer of randomly chosen single-qubit Pauli rotations and a fixed sequence of CZ gates. This construction aims to approximate a unitary $t$-design, introducing sufficient randomness for statistical benchmarking.
A general expression for \( U(\boldsymbol{\theta}) \) in the RPA can be written as
\begin{align}
U(\bm{\theta}) = \prod_{\ell=1}^{L} \left( U_{\text{ent}}^{(\ell)} \cdot U_{\text{ran}}^{(\ell)}(\boldsymbol{\theta}^{(\ell)}) \right) \cdot
\left( \bigotimes_{j=1}^N R_Y^{(j)}\left(\frac{\pi}{4}\right) \right),
%\left( \bigotimes_{j=1}^N R_{P_j}^{(j)}(\theta_j) \right) \left( \bigotimes_{j=1}^N R_Y^{(j)}\left(\frac{\pi}{4}\right) \right),
\end{align}
where $U_{\text{ran}}^{(\ell)}(\boldsymbol{\theta}^{(\ell)}) = 
\bigotimes_{j=1}^N R_{P_j}(\theta_j^{(\ell)})$
and \( P_j \in \{X, Y, Z\} \) is a randomly selected Pauli operator for each qubit \( j \), i.e.,  \( R_{P_j}(\theta_j) = \exp\left(-i \frac{\theta_j}{2} P_j \right) \) is a rotation around the Pauli axis \( P_j \) by angle \( \theta_j \).
The total number of parameters is $d = NL$.
This ansatz is shallow but introduces sufficient randomness to study expressibility and trainability, while allowing analytical studies on gradient statistics and BPs.

\subsection{Cost function}
In VQAs, the cost function is commonly written as \cite{Cerezo2021}
\begin{align}
C(\bm{\theta}) = \sum_k f_k \Big(\, \mathrm{Tr}\big[O_k \, U(\bm{\theta}) \rho_k U^\dagger(\bm{\theta})\big]\Big),
\end{align}
where $\rho_k$ denote input quantum states (such as data-encoded states or initial states), $O_k$ are Hermitian observables representing measurement operators or problem-specific Hamiltonians, and $f_k$ are functions applied to the expectation values. This formulation is sufficiently general to encompass a wide range of VQAs, including those for optimization, simulation, and quantum machine learning, by appropriately choosing $\rho_k$, $O_k$, and $f_k$.

For instance, in VQEs, $O_k$ correspond to the system Hamiltonian $H$, which can be written as a linear combination
\(
H = \sum_k c_k H_k,
\)
where each $H_k$ is a tensor product of Pauli operators and $c_k$ are real coefficients.

For benchmarking, we optimize the expectation value of a given Hamiltonian, such as \( H = Z_1 Z_2 \), where \( Z_j \) denotes the Pauli-\( Z \) operator acting on the \( j \)-th qubit \cite{McClean2018}. 
Although this Hamiltonian does not represent a computationally hard problem, it serves the purpose of providing a VQE that has BPs, which allows us to analyze the statistical behavior of the gradient and explore ways to mitigate the BPs in the VQE.
The cost function, which guides the optimization process, is defined as
\begin{align}
        C(\boldsymbol{\theta}) = \langle \boldsymbol{0}| U(\boldsymbol{\theta})^\dagger H U(\boldsymbol{\theta}) | \boldsymbol{0} \rangle,
\end{align}
where $|\bm 0\rangle = |00\cdots0\rangle$.
This cost function measures the expectation value of the Hamiltonian \( H \) under the 
evolved state   \( U(\boldsymbol{\theta})|\bm 0\rangle \). By adjusting \( \boldsymbol{\theta} \), the algorithm seeks to minimize \( C(\boldsymbol{\theta}) \), effectively finding the optimal quantum state which can be considered an approximation for the ground state of $H$.

\subsection{BPs in one direction}
Figure~\ref{fig:5} provides numerical evidence of BPs for two ans\"atze that represent HEA and RPA at a depth $L = 20$, focusing on the $\theta_1$ direction. In both cases, the cost function $C(\bm{\theta})$ is plotted as a function of $\theta_1$, with all other parameters fixed at randomly chosen values [Fig.~\ref{fig:5}(a, b)]. This allows us to visualize the one-dimensional cost function landscape along the $\theta_1$ direction, consistent with gradient calculations where other parameters are held constant.

For small qubit numbers $N$, the cost function landscape $C(\bm{\theta})$ shows clear variation with respect to $\theta_1$, enabling effective gradient-based optimization. As $N$ increases, the landscape becomes increasingly flat, indicating the evolution towards everywhere-flat BPs.
Importantly, no localized dip structure develops as we gradually increase $N$.

Figures~\ref{fig:5}(c, d) show the statistical distributions of $\partial_{\theta_1} C$, obtained by randomly sampling all parameters in $\bm{\theta}$ over $[0, 2\pi]$. We generate 100 sample for each $N$. 
We observed that the derivative distribution decreases overall while maintaining a nearly constant spread, similar to the inset of Fig.\ref{fig:1}(g), where the everywhere-flat BPs exist. The variance $\text{Var}[\partial_{\theta_1} C]$ exhibits exponential decay with slopes of $-0.74$ (HEA) and $-1.15$ (RPA). 
In contrast, the first region of Fig.\ref{fig:1}(g) shows the largest derivatives remaining large, while the smallest shrink significantly. This comparison supports the presence of everywhere-flat barren plateaus in the VQE.

The BPs arise from the concentration of measure in high-dimensional Hilbert spaces, where randomly parameterized circuits yield nearly uniform cost function distributions. Consequently, the cost function becomes insensitive to individual parameter changes, and gradients vanish exponentially with system size. Despite structural differences, both HEA and RPA possess sufficient expressivity and randomness to exhibit this effect even at fixed $L$.

Furthermore, the probability $\Pr(|\partial_{\theta_1} C| \geq \delta)$ with a threshold $\delta = 10^{-1}$ drops to zero at $N = 7$ and $N = 5$ for two ans\"atze HEA and RPA, respectively. 
The specific choice of threshold $\delta = 10^{-1}$ is not intrinsically significant but is chosen to match the gradient scale in our simulations, enabling consistent and meaningful comparisons across ans\"atze. %The probability $\Pr(|\partial_{\theta_1} C| \geq \delta)$ serves as a direct and interpretable measure of trainability. 
Using a fixed threshold across different ans\"atze provides a concrete basis for comparison: a faster decay of this probability reveals an earlier emergence of flat regions in the cost function landscape, indicating a more pronounced BPs.

\subsection{Comparison between different directions in parameter space}
\begin{figure}[t]
\centering
\includegraphics[width=\linewidth]{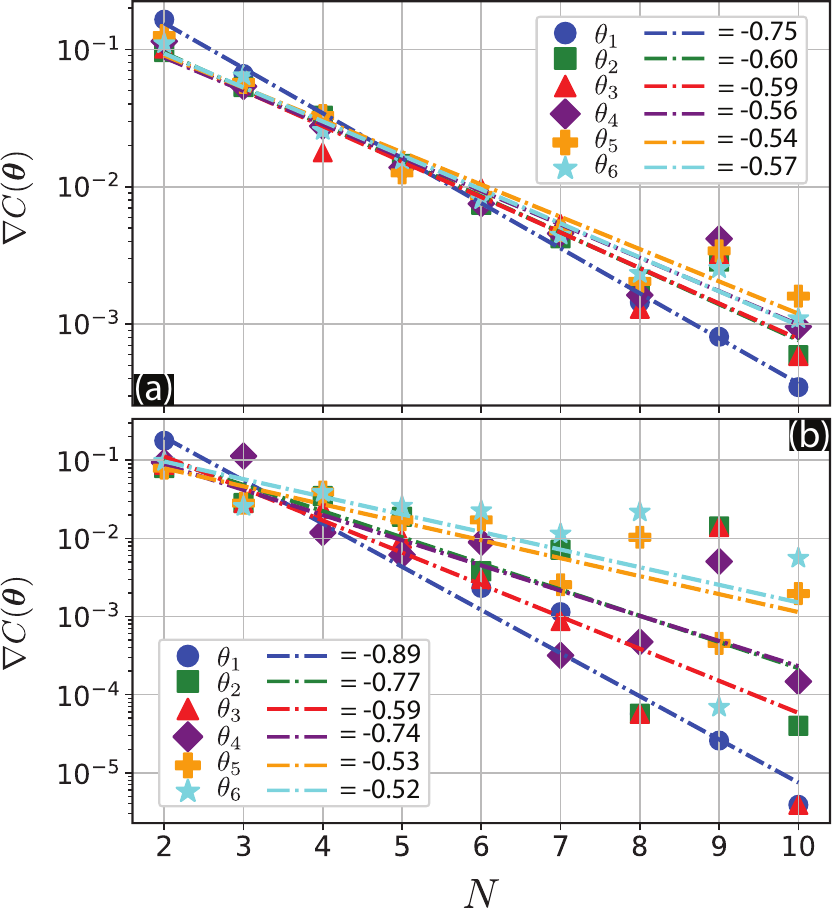}
\caption{Absolute values of the first 6 gradients as $\nabla C(\boldsymbol{\theta}) = \big[|\partial_{\theta_1} C(\boldsymbol{\theta})|, \dots, |\partial_{\theta_6} C(\boldsymbol{\theta})|\big]$ as  functions of qubit number $N$ for (a) HEA and (b) RPA.}
\label{fig:6}
\end{figure}

We next examine BPs along others parameter directions by analyzing the first six parameters, $\theta_1$ to $\theta_6$. The results are shown in Fig.~\ref{fig:6}. For each $\theta_k$, we compute the variance $\text{Var}[\partial_{\theta_k} C]$ over random parameter samples. Exponential decay is observed across all directions, not only for $\theta_1$, indicating that the flatness of the cost landscape extends throughout the parameter space.

The slopes vary depending on the parameters. For example, in the HEA in Fig.~\ref{fig:6}(a), the slopes range from approximately $-0.54$ to $-0.75$, while for RPA in Fig.~\ref{fig:6}(b), they range from $-0.52$ to $-0.89$. These differences reflect the non-uniform structure of the ans\"atze but consistently confirm the presence of BPs in all directions.

\subsection{The effect of circuit depth}

\begin{figure}[t]
\centering
\includegraphics[width=\linewidth]{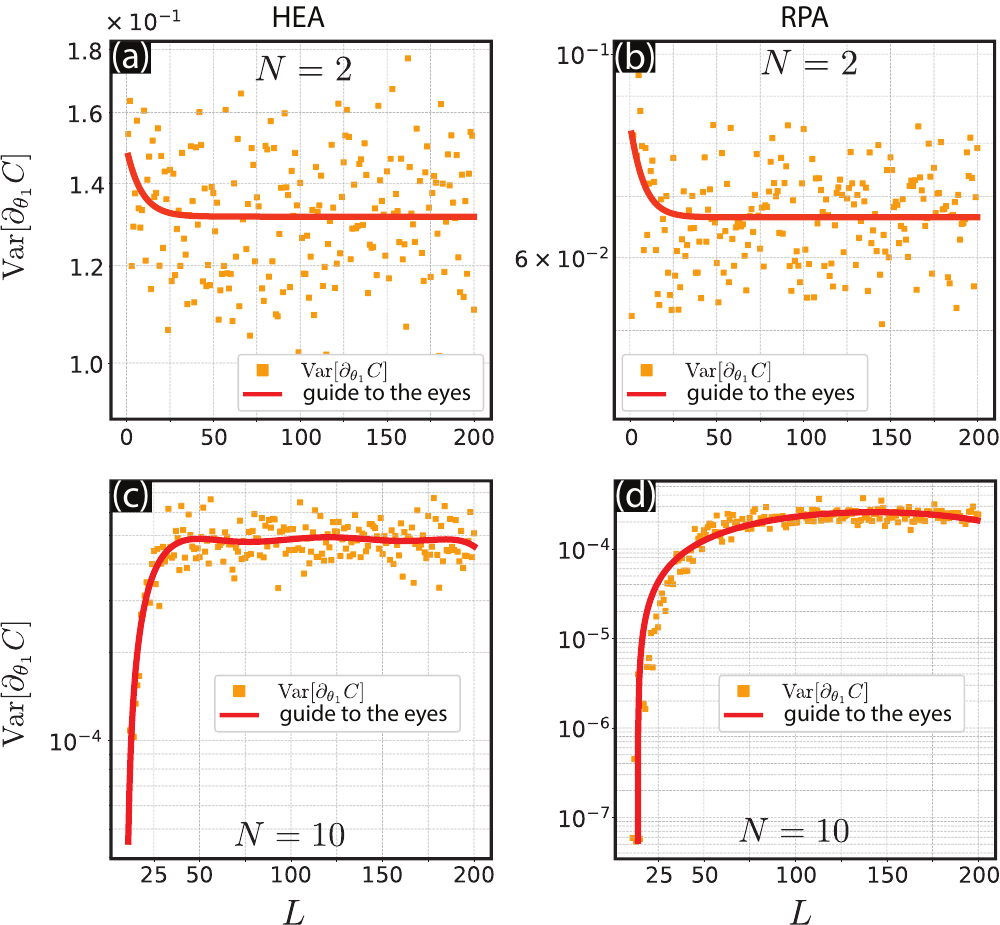}
\caption{Variance of the cost function derivative ${\rm Var}[\partial_{\theta_1} C]$ as a function of the circuit depths $L$. Panels (a) and (b) show results for $N = 2$ qubits using the HEA and RPA ans\"atze, respectively. In both cases, the variance decreases with increasing $L$, approaching a constant value consistent with the behavior of approximate 2-designs. Panels (c) and (d) display the corresponding results for $N = 10$, where the variance instead increases with $L$ and saturates at an upper bound, indicating a different scaling regime.
}
\label{fig:7}
\end{figure}

We now explore how the number of layers $L$ affects the emergence of BPs. The results are shown in Fig.~\ref{fig:7}. Panels (a) and (b) depict the variance ${\rm Var}[\partial_{\theta_1} C]$ as a function of $\theta_1$ for different values of $L$, using the HEA and RPA ans\"atze respectively, with system size $N = 2$. In both cases, the variance decreases with increasing $L$, eventually stabilizing at a limiting value. This behavior is consistent with convergence to an exact 2-design, particularly in the RPA case, where the circuit becomes highly expressive at large depths.

In contrast, for $N = 10$, shown in panels (c) and (d), the trend reverses: the variance increases with $L$ and then saturates. Note that the saturation value is smaller than the one in the $N = 2$ case. 

These findings are consistent with previous theoretical studies on the large-$L$ regime, where the variance tends to a constant that decreases with increasing $N$ \cite{McClean2018}. However, we observe a difference at small depths. Specifically, for $N = 10$, the variance remains low when $L$ is small, indicating the presence of BPs in the shallow circuit regime. This highlights that BPs do not develop uniformly across all circuit depths and system sizes. For example, shallow circuits in large systems can maintain negligible gradients.

This behavior can be understood through the statistical concentration of gradients. As established in \cite{McClean2018}, deeper circuits increasingly resemble random unitary ensembles, eventually forming approximate 2-designs. This leads to gradients approaching zero, characteristic of BPs. However, at shallow depths, the circuit remains far from fully random. Its limited expressivity preserves structure and correlations among parameters, preventing the complete vanishing of gradients.

In our case, when $L$ is small, the circuit lacks sufficient depth to induce full randomization, and gradient distributions remain broad. As $L$ increases, the circuit becomes more expressive, leading to more randomized transformations. This increases the likelihood of gradient suppression due to averaging effects.

%Figures (c) and (d) show the corresponding derivative distributions $\partial_{\theta_1}C$, along with the probability \(\mathrm{Pr}(|\partial_{\theta_1} C| \geq\delta)\) and the variance ${\rm Var}[\partial_{\theta_1}C]$. In (c), for \( L = 3 \), the derivatives are concentrated around \( 10^{-14} \), the probability at \(\delta = 10^{-6}\) drops to zero, and the variance is on the order of \(10^{-26}\), indicating a BP. For \( L \ge 4 \), gradients increase to the order of \(10^{-1}\), the probability reaches 1, and the variance is much larger, indicating no BP. A similar trend holds for RPA: barren plateaus are present for \( L < 5 \) but vanish when \( L \ge 5 \).
%The variance is also approach the exact 2-design ansatz.

\begin{figure}[t]
\centering
\includegraphics[width=\linewidth]{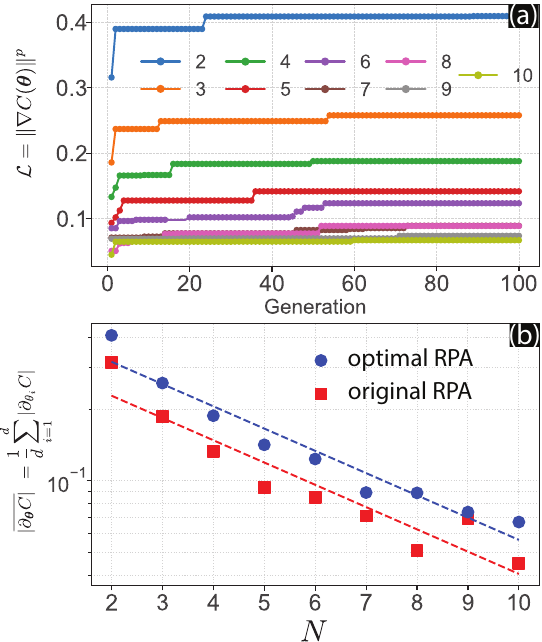}
\caption{Numerical analysis of the GA optimization applied to RPAs. (a) Fitness function versus the number of generation for system sizes $N = 2$ to $N = 10$. Initially low fitness values, especially for large $N$ reflect the poor trainability of randomly initialized circuits. As the GA evolves, fitness improves and saturates, indicating successful optimization. (b) Average gradient cost function $\overline{\big|\partial_{\bm\theta} C\big|}$ as a function of $N$, comparing original (unoptimized) RPAs with GA-optimized RPAs. The optimized circuits exhibit significantly higher gradients across all $N$, confirming improved trainability and suppression of barren plateaus.}
\label{fig:8}
\end{figure}

\section{Barren plateau mitigation}\label{sec4}
In this section, we turn to a somewhat different subject. We propose a simple strategy to mitigate the deleterious effects of BPs using a genetic algorithm (GA) \cite{Hai_2024}. The GA is designed to optimize the randomness of the ansatz. For definiteness, we focus on the RPA and
aim to increase the magnitude of its gradient, pushing it away from the zero-gradient regime.

The GA is designed to evolve RPAs into ones that yield large, non-zero gradients. We define the fitness function to an ${\rm L}^p$ norm objective of the gradient vector $\nabla C(\boldsymbol{\theta})$ as
%$$
%\mathcal{L} = | \partial_{\theta_1} \mathcal{C} |,
%$$
\begin{align}
\mathcal{L} = \|\nabla C(\boldsymbol{\theta})\|^p 
= \Bigg(\sum_{i=1}^d \big| \partial_{\theta_i} C \big|^p\Bigg)^{1/p}, \\
\text{s.t.} \quad \Pr\big( \big|\partial_{\theta_i} C\big| \geq \varepsilon \big) = 1. \label{eq:con9}
\end{align}
where $d$ is the number of parameters, and $\varepsilon > 0$ is a threshold, which can be chosen based on the statistical results of the original ansatz.
The condition ${\rm Pr}(| \partial_{\theta_i} C| \geq \varepsilon) = 1$ ensures that the gradients for all the parameters in the optimized ans\"atze will non-vanish.
The fitness function $\mathcal{L}$ allows us to compare various generated ans\"atze and choose the best one.

The GA proceeds as follows: we first generate a set of ans\"atze. We then evaluate their fitness using the defined fitness function $\mathcal{L}$. The two highest-scoring  ans\"atze are selected and combined through crossover to produce a new candidate. 
Crossover refers to exchanging subsets of parameterized gates or circuit layers between two parent circuits to construct a new variational ansatz that inherits structural features from both.
A small mutation is then applied, such as, a random Pauli gate in the circuit is replaced by another Pauli gate with a low but finite probability. The fitness of the new ansatz is evaluated, and the evolution process is repeated either for a fixed number of generations or until the condition \eqref{eq:con9} satisfy with a certain statistical number of each parameter.

By using the GA optimization, the final ansatz will develop a gradient structure that is non vanishing. 
During the GA process, the parameters $\boldsymbol{\theta}$ are generated randomly, since the objective is to optimize the structure of the ansatz, not the parameter values themselves.

Figure~\ref{fig:8} presents the results of the GA optimization. Fig.~\ref{fig:8} (a) shows the fitness as a function of the number of generations for system sizes $N = 2$ to $N = 10$ and $p = 1$, $\epsilon = 0.05$. Initially, the RPA circuits are randomly generated, leading to low fitness values, especially for larger $N$, due to poor gradient flow and a higher likelihood of BPs. As the GA evolves, the fitness improves steadily and eventually saturates, indicating convergence toward more trainable circuit structures.

Correspondingly, Fig.~\ref{fig:8} (b) shows the average gradient cost function, i.e., $\overline{\big|\partial_{\bm\theta} C\big|}= \frac{1}{d}\sum_{i = 1}^d \big|\partial_{\theta_i}C\big|$, plotted against the system size $N$. We compare two cases: the original (unoptimized) RPA and the optimized RPA obtained via the GA. A clear enhancement in the average gradient is observed for the optimized RPA across all values of $N$. This improvement reflects the GA’s ability to navigate the parameter space and identify circuit architectures that preserve gradient magnitude, thereby mitigating the impact of BPs.

The original case reflects the tendency of randomly initialized deep parameterized quantum circuits to approximate 2-designs, resulting in the emergence of BPs \cite{McClean2018}. The GA-based optimization case effectively reshapes the structure of the RPA in the setting where it optimizes and selects only the best candidate among various randomly generated ans\"atze.
As a result, the optimization process actively resists the BPs, especially in the case of larger qubits number. The observed increase in the average $\overline{\big|\partial_{\bm\theta} C\big|}$ confirms the enhancement in trainability and supports the effectiveness of the proposed method.

\section{Conclusion}\label{sec5}
We developed a statistical framework to analyze BPs in VQAs. By studying an analytically tractable cost function in Gaussian functions, we identified three distinct types of BPs. The first arises near sharp dips or critical points in the parameter space and the second has a sharp gorge.  Gradients in these cases are small but still informative, allowing optimization algorithms to make progress. The third, more detrimental type, occurs in uniformly flat regions of the landscape, where gradients vanish across broad domains, significantly impeding trainability.

We then apply the framework to HEA and RPA ans\"atze in VQE settings, where we found that only the second type of BP appears. To mitigate the BPs in VQE, we proposed a GA-based optimization scheme aimed at reshaping the cost landscape. By selectively enhancing structure and gradient visibility, our method effectively suppresses the occurrence of flat barren plateaus and enables more efficient training, even in larger systems.
These findings contribute a new perspective on the barren plateau problem and suggest a practical pathway toward improving the scalability of VQAs. 

\begin{acknowledgments}
This paper is supported by JSPS KAKENHI Grant
No.23K13025.
SA was supported by Japan’s Ministry of Education, Culture,
Sports, Science and Technology (MEXT) Quantum Leap Flagship Program Grant Number
JPMXS0120319794.
\end{acknowledgments}
%\newpage
%\appendix
%\section{}

\bibliography{refs}

\end{document}